\documentclass[sigconf]{acmart}
\AtBeginDocument{%
  \providecommand\BibTeX{{%
    \normalfont B\kern-0.5em{\scshape i\kern-0.25em b}\kern-0.8em\TeX}}}

\usepackage{graphicx}
\usepackage{subcaption}
\usepackage{todonotes}

\copyrightyear{2023}
\acmYear{2023}
\setcopyright{rightsretained}
\acmConference[CBMI 2023]{International Conference on Content-based Multimedia Indexing}{September 20--22, 2023}{Orleans, France}
\acmBooktitle{International Conference on Content-based Multimedia Indexing (CBMI 2023), September 20--22, 2023, Orleans, France}
\acmDOI{}
\acmISBN{}

\begin{document}
\title{Memories in the Making: Predicting Video Memorability with Encoding Phase EEG}

\author{ Lorin Sweeney, Graham Healy and Alan F. Smeaton}
%\orcid{}
%\author{}
%\orcid{}
%\author{}
\email{lorin.sweeney8@mail.dcu.ie}
%\orcid{0000-0003-1028-8389}

\affiliation{%
  \institution{Insight SFI Research Centre for Data Analytics,\\Dublin City University, Glasnevin, Dublin 9}
  \country{Ireland}
}
\renewcommand{\shortauthors}{Sweeney et al.}

\begin{abstract}
In a world of ephemeral moments, our brain diligently sieves through a cascade of experiences, like a skilled gold prospector searching for precious nuggets amidst the river's relentless flow. This study delves into the elusive "moment of memorability"---a fleeting, yet vital instant where experiences are prioritised for consolidation in our memory. By transforming subjects' encoding phase electroencephalography (EEG) signals into the visual domain using scaleograms and leveraging deep learning techniques, we investigate the neural signatures that underpin this moment, with the aim of predicting subject-specific recognition of video. Our findings not only support the involvement of theta band (4-8Hz) oscillations over the right temporal lobe in the encoding of declarative memory, but also support the existence of a distinct moment of memorability, akin to the gold nuggets that define our personal river of experiences.
\end{abstract}





%  \caption{Something cool and relevant.}
%  \Description{}

\maketitle

\section{Introduction}
In a world awash with fleeting moments, the river of time relentlessly splashes us with the experience of being. Amidst this mercurial torrent of sensory droplets---each vying for a place in the precious annals of our memory---our brain stands as the vigilant gatekeeper, painstakingly managing the flow of water and deciding which droplets will reach the reservoir of our memory. However, our reservoir---like any storage system---is subject to constraints of capacity and encoding efficiency. We accordingly posit that a critical moment of memorability should exist, an ephemeral yet potent point in time which captures the essence of an experience, and assigns it a ``remembering priority'', which will ultimately determine its fate within the annals of our memory when consolidation comes around.

In this paper, we explore the predictive power of encoding phase electroencephalography (EEG) signals, recorded from subjects during video stimulus presentation, to predict subject specific recognition upon subsequent (24--72 hours later) re-viewing. We transform the EEG signals into the visual domain by turning them into scaleograms with a continuous wavelet function, which allows us to avail of state-of-the-art visual deep learning techniques. By leveraging  temporal and spatial information contained within the EEG data, we position ourselves to capture the moment of memorability---a moment of encoding that corresponds to a remembering moment. We hypothesise that the neural signals recorded during this moment of memorability will differ from those recorded during forgettable moments, and that these differences can be used to predict whether a given subject will remember a given video. We employed a two factor study design---comparing subject-independent (SI) and subject-dependent (SD) training approaches, and comparing single electrode and composite 28 electrode scaleogram images---in order to evaluate the generalisability of our approach and whether theta band (4--8Hz) activity over the right temporal lobe (channel P8 being the closest available electrode location to this brain region in this dataset), which has been implicated in memory formation \cite{osipova2006theta}, leads to more accurate predictions. 

 \section{Related Work}
Although the nature and constitution of people’s memories remains elusive, and our understanding of what makes one thing more/less memorable than another is still nascent, combining computational (e.g., machine learning) and neurophysiological (e.g., electroencephalography; EEG) tools to investigate the mechanisms of memory has the potential to offer insights that would be otherwise unobtainable. While EEG is not a tool that can directly explain the factors that make an experience more/less memorable, it can help us trim the umbral undergrowth surrounding the subject, and offer a potential leap forward in our understanding of the interplay between the mechanisms of memory and memorability. 

The use of EEG has proven to be an effective tool in the investigation of the neural mechanisms that underpin memory formation and recall, providing insight into the timing and general location of these processes \cite{sanquist1980electrocortical, karis1984p300, klimesch1999eeg, noh2014using}. Even though the application of machine learning to EEG is an active area of interest---enabling the classification of various cognitive states and processes, such as emotion \cite{wang2014emotional}; mental tasks, e.g., relaxation, counting, multiplication, word generation \cite{liang2006classification, lin2009classification}; sleep stages \cite{ebrahimi2008automatic}; and the automation, or augmentation of neurological diagnostics \cite{ieracitano2020novel, lehmann2007application, hosseinifard2013classifying, engemann2018robust}---the use of EEG to predict visual memorability has been limited to static stimuli such as images \cite{jo2020prediction}, leaving video entirely unexplored.

\section{Dataset \& Methodology }

EEGMem \cite{ME2022} is an extension of the Memento10k \cite{mem10k} video memorability dataset. It is comprised of encoding phase EEG recordings captured from 12 participants as they watched a subset of the Memento10k videos, and was collected as part of the 2022 MediaEval Predicting Video Memorability task \cite{ME2022}. EEGMem data collection was split into two phases: an encoding phase, during which participant EEG was recorded while watching a continuous stream of 1,000 videos, and an online recognition phase, which took place 24 to 72 hours after the encoding phase, and during which they re-watched those 1,000 videos mixed with an additional 1,000 unseen Memento10k videos, responding with the keyboard press if they recognised a video from the encoding phase, thus providing binary annotations.

\subsection{Pre-processing}

The raw EEG data was first referenced with a common average, and band-pass filtered between 0.1--30Hz (N.B., this does not affect the time-frequency resolution of the EEG signal, which was sampled at 1000Hz). Independent Component Analysis (ICA) was then used to remove artefacts and reject trials using subject-specific thresholds. Binary annotations from the recognition phase were used to label subject specific encoding phase EEG trials and their associated Memento10k videos as ``True'' for correctly remembered videos (true positive), and ``False'' for all other outcomes.

\subsection{Feature-extraction}

\begin{figure}[ht]
    \centering
    \begin{minipage}[b]{0.23\textwidth}
        \includegraphics[width=\textwidth]{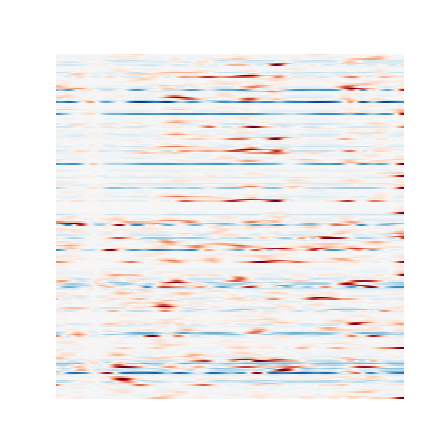}
        \caption*{(a) 28-channel composite}
    \end{minipage}
    \hfill
    \begin{minipage}[b]{0.23\textwidth}
        \includegraphics[width=\textwidth]{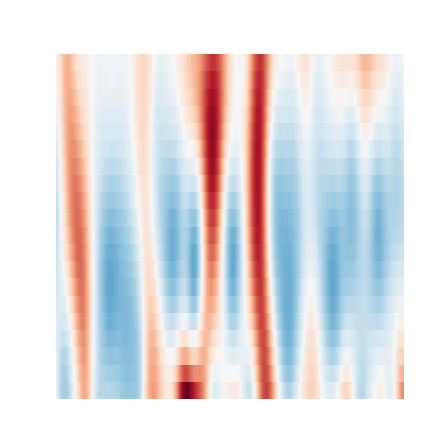}
        \caption*{(b) single channel}
    \end{minipage}
    \caption{Examples of the two types of scaleogram images.}
    \label{fig:scaleograms}
\end{figure}

With the goal of leveraging the state-of-the-art Vision Transformer (ViT) architecture, we turned the filtered EEG data into one of two kinds of scaleogram with a Morlet wavelet function---one with all 28 electrode channels, and the other with a single channel (Figure~\ref{fig:scaleograms}). The choice of Morlet wavelet scaleograms was motivated by its superior time-frequency resolution, and our desire evaluate the efficacy of the theta band of frequencies (4--8Hz) over the right temporal lobe (electrode P8) as a predictive feature of video memorability. Single channel scaleograms were generated with 20 linearly spaced frequencies from 4 to 8Hz, while each channel in the 28 channel scaleograms were generated with 8 linearly spaced frequencies from 3 to 17Hz. The number of cycles used for both sets of frequencies were determined by a logarithmic function that balances the time-frequency resolution trade-off. For both scaleogram types we use a baseline period of -0.25 to 0 seconds and apply z-score normalisation, and then average across trials and visualised for the time period -0.5 to 3 seconds (the full duration of the video). The final 28 channel scaleogram image is a 4 by 7 grid of individual scaleograms, one for each channel. 
 
\subsection{Fine-tuning ViT Models}

The ViT \cite{dosovitskiy2020image} architecture was chosen due to its lower image-specific inductive bias and its current state-of-the-art status in image recognition tasks. We fine-tuned (i.e., no layers were frozen, a new task specific head was created, and the model is re-trained) a total of 52 distinct ViT-large models pre-trained on the ImageNet-21k dataset. Each model was fine-tuned on one of four types of data: 28 Channel, Fp1, P8, and Frames (using the first, middle, and last frame). The fine-tuning procedure was executed under two distinct training categories: Subject-Independent (SI) and Subject-Dependent (SD), with 12 SI models (one per subject), and 4 SD models trained for each data type. SI models were fine-tuned on all subjects aside from one, in a leave-one-out fashion, and tested on a stratified test set consisting of 15\% of the left out subject's data. SD models were fine-tuned on all subject data aside from a combined test set composed of each subject's stratified test set. All models were trained using the AdamW optimiser with a learning rate of 2e-5, a cosine learning rate scheduler with a 0.1 weight decay, and both early stopping and a dropout rate of 0.65 were used to mitigate overfitting.

\section{Results}
\begin{table*}[t]
\caption{AUC ROC scores for all models. (Subject Mean excludes Frames column data).}
\label{tab:results1}
\hspace*{-0.2cm}
\begin{tabular}{cccccccccccccc}
    \toprule
     &\multicolumn{2}{c}{28 Channel} && \multicolumn{2} {c}{Fp1} && \multicolumn{2} {c}{P8} && \multicolumn{2} {c}{Frames} \\
    \cline{2-3}\cline{5-6}\cline{8-9}\cline{11-12}
    SID & SI & SD     && SI & SD && SI & SD && SI & SD &&  Subject Mean\\
    \midrule
    S-2 & 0.445 & 0.540 && 0.488 & 0.509 && 0.558 & \textbf{0.572} && 0.487 & 0.477 && 0.518 \\
    S-4 & 0.482 & 0.466 && 0.483 & 0.470 && \textbf{0.634} & 0.577 && 0.472 & 0.496 && 0.518 \\
    S-9 & 0.437 & 0.530 && 0.459 & 0.478&& 0.569 & \textbf{0.613}  && 0.498 & 0.501 && 0.514 \\
    S-10 & 0.462 & 0.457&& 0.489 & 0.502  && 0.568 & \textbf{0.578} && 0.515 & 0.483 && 0.509\\
    S-13 & 0.430 & 0.474 && 0.464 & 0.508 && 0.574 & \textbf{0.596} && 0.492 & 0.500 && 0.508\\
    S-16 & 0.383 & 0.523 && 0.497 & 0.528 && 0.534 & \textbf{0.564} && 0.489 & 0.482 && 0.503\\
    S-19 & 0.393 & 0.405 && 0.509 & 0.516 && 0.578 & \textbf{0.636}  && 0.514 & 0.488 && 0.506\\
    S-30 & 0.388 & \textbf{0.441} && 0.331 & 0.349 && 0.340 & 0.436 && 0.502 & 0.492 && 0.381\\
    S-31 & 0.626 & \textbf{0.655} && 0.521 & 0.556 && 0.510 & 0.520 && 0.479 & 0.506 && 0.565\\
    S-36 & 0.696 & 0.540 && 0.599 & 0.532 && \textbf{0.707} & 0.564 && 0.511 & 0.481 && \textbf{0.606}\\
    S-37 & 0.562 & 0.323 && 0.531 & 0.401 && 0.614 & \textbf{0.632} && 0.503 & 0.486 && 0.510\\
    S-41 & 0.566 & 0.386 && 0.501 & 0.343 && \textbf{0.567} & 0.538  && 0.500 & 0.506 && 0.484\\
    \midrule
    Mean & 0.489 & 0.478 && 0.489 & 0.474 && 0.563 & \textbf{0.567} && 0.497 & 0.492 && \\
    \bottomrule
\end{tabular}
\end{table*}

\begin{figure}
    \centering
    \includegraphics[clip,trim=0 0 0 0, width=0.48\textwidth]{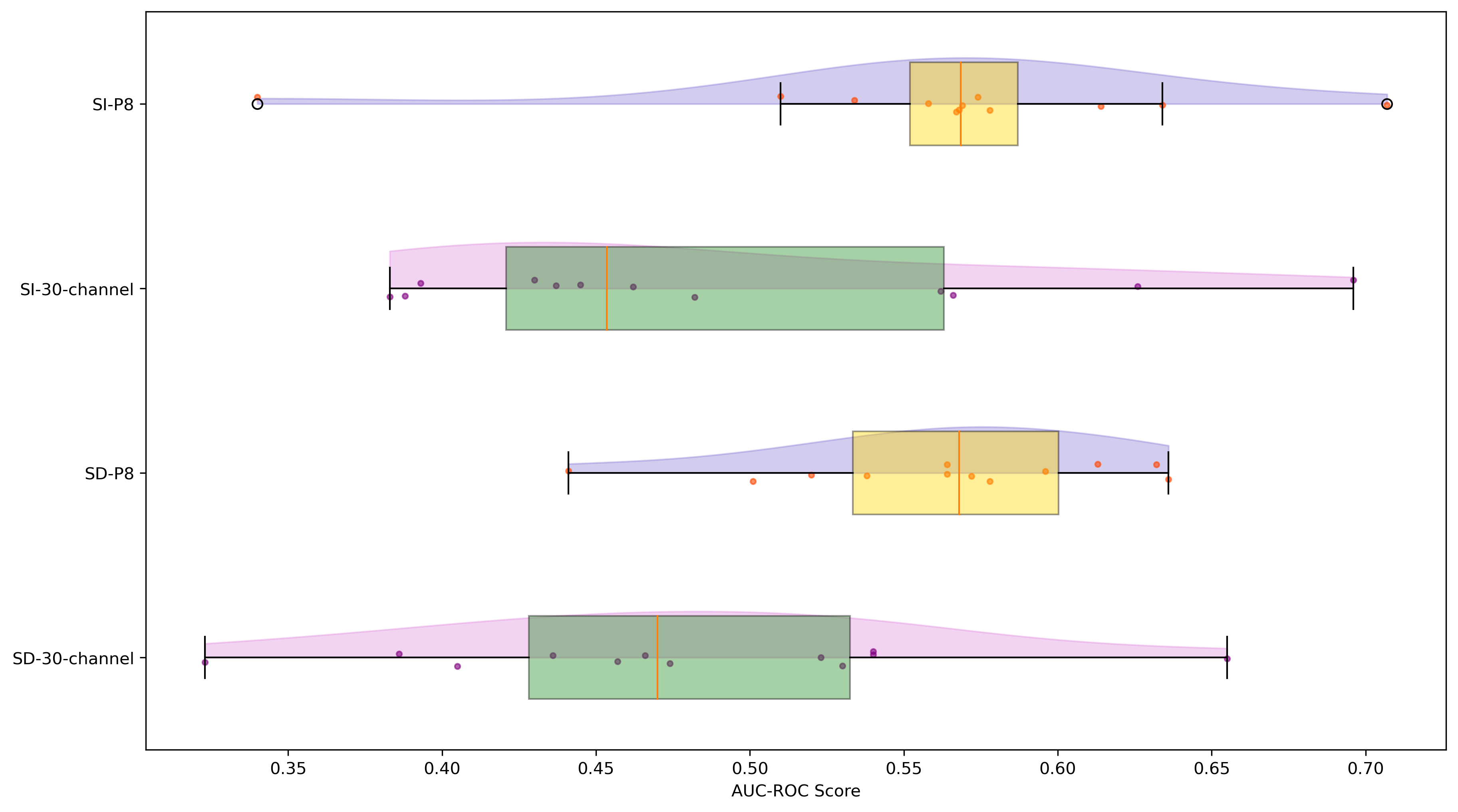}
    \caption{Raincloud plot comparing 28 channel \& P8 channel scaleogram trained models.}
    \label{fig:rain}
\end{figure}

\begin{figure}
    \centering
    \includegraphics[clip,trim=0 0 0 0, width=0.48\textwidth]{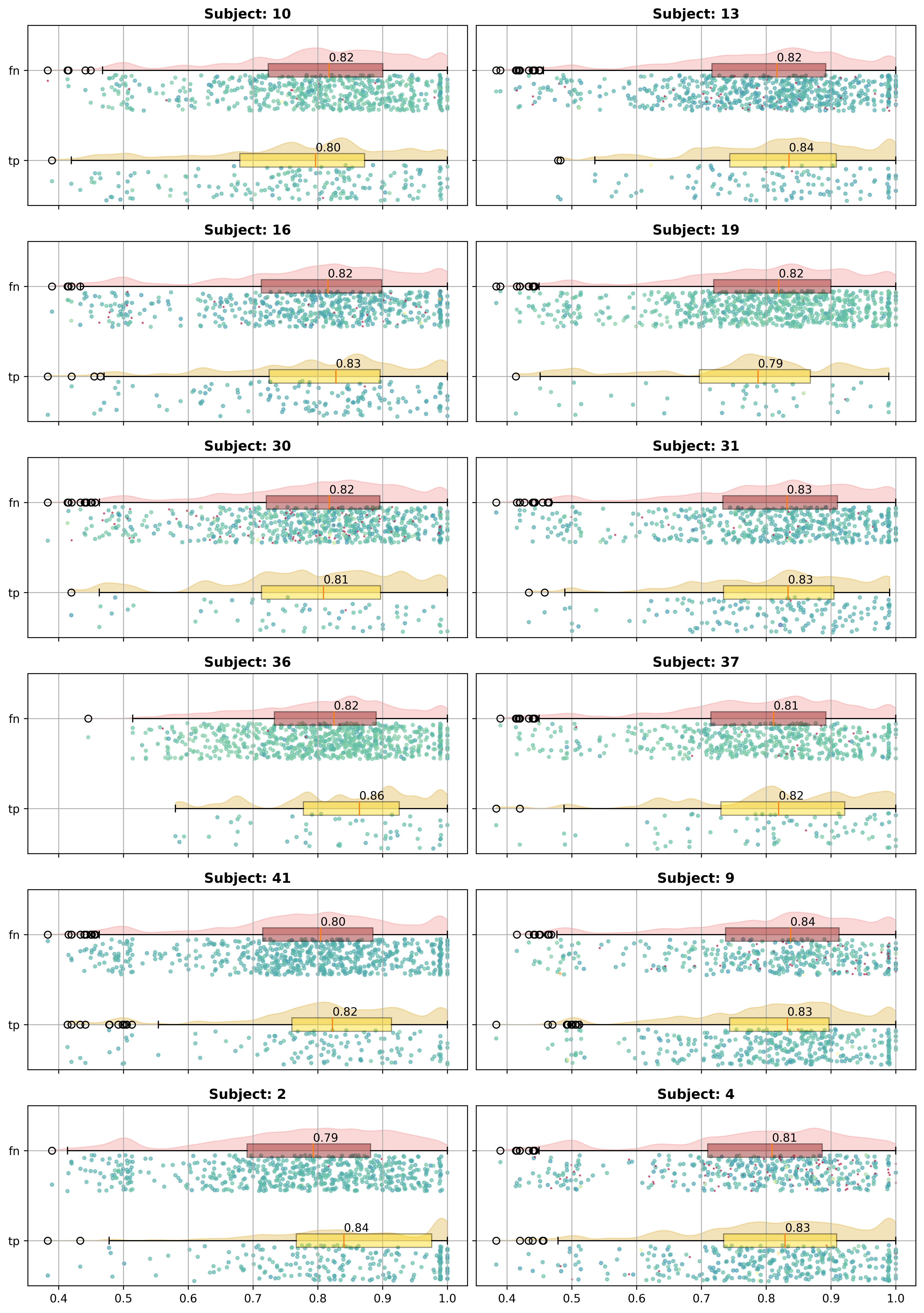}
    \caption{Raincloud plots for each subject plotting the difference in ground-truth memorability scores (x-axis) for remembred (tp) and forgotten (tn) videos. The colour and size of each dot is proportional to the subjects response reaction time when viewing the video for the second time.}
    \label{fig:rain2}
\end{figure}

Model performance was evaluated using the Area Under the Receiver Operating Characteristic Curve (AUC-ROC) as there were only two prediction classes---remembered and forgotten. Individual classification results for each subject (column headed SID) are displayed in Table~\ref{tab:results1}, for both the SI and SD models. This shows that our ViT models trained on channel P8 scaleograms are the best performing, with a mean AUC-ROC of 0.567 for SD trained models, and a mean AUC-ROC of 0.563 for SI trained models. A set of ViT models were trained on a random channel's (Fp1) scaleograms in order to account for the possibility that data resolution was the factor driving the 28 channel and P8 single channel performance difference, which the results suggest was not the case. An additional set of ViT models (Frames) were trained, which were trained on video frame data rather than EEG data, the results of which were random, with no discernible difference between SI and SD training, nor notable difference in subject performance, which is logically coherent with the fact that video frame data is not influenced by the subject viewing it. 

\begin{figure}
    \centering
    \includegraphics[clip,trim=0 0 0 0, width=0.48\textwidth]{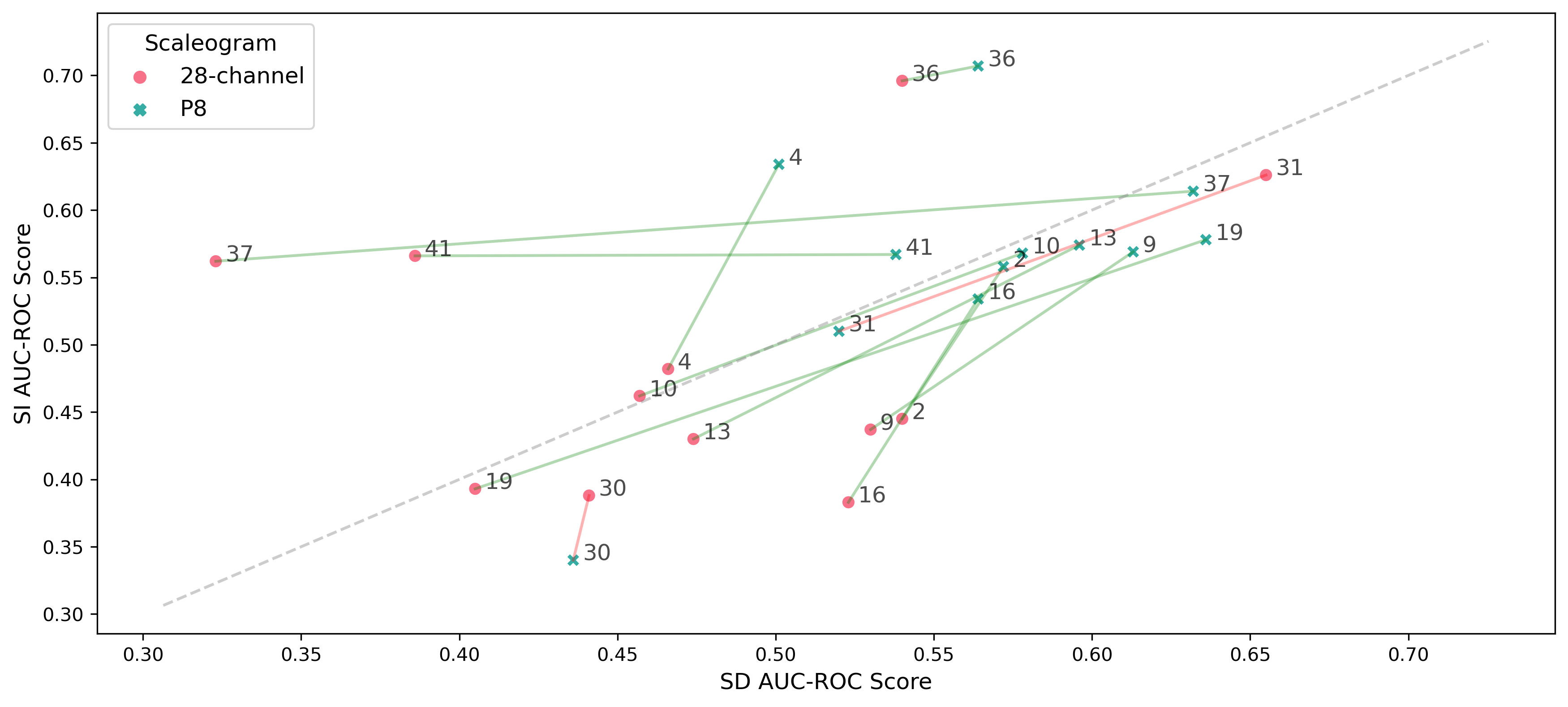}
    \caption{AUC-ROC scatter plot for scaleogram comparison. Subject specific P8 vs 28 Channel model performance difference is illustrated with coloured lines---green indicating P8 improvement, and red indicating 28 Channel improvement.}
    \label{fig:scatter}
\end{figure}

Our repeated-measures design (each participant is in both subject-dependent and subject-independent conditions, and their data contribute to both scaleogram types), allowed us to perform a paired t-test, finding that the AUC-ROC scores for models trained with P8 channel scaleogram images are significantly (t = 3.243, p = 0.0036) greater than those trained with 28 channel scaleogram images, adding weight the hypothesis that theta band (4--8Hz) oscillations over the right temporal lobe (channel P8), predict encoding of declarative memory \cite{osipova2006theta}. Figure ~\ref{fig:rain} and Figure ~\ref{fig:scatter} provide visual illustrations of the difference between P8 and 28 channel trained scaleogram images. Although we did not find a significant interaction between SD and SI training, we can see that more points in Figure ~\ref{fig:rain} lie below the line of best fit, indicating that more models trained on SD performed better than their SI trained counterparts, which makes sense from a training perspective, as EEG data can be highly subject specific, and not having any subject training examples can hinder prediction performance. Figure~\ref{fig:rain} additionally highlights the impact of 28 channel vs channel P8 scalograms on a per subject basis, with a green line indicating that either SD or SI AUC ROC scores improved and red lines indicating that they both worsened.

Subjects S-30 and S-36 sit on two ends of the performance spectrum, with S-30 producing the worst performances across the board (an average AUC ROC of 0.381), and S-36 producing the highest average AUC ROC of 0.606 and the highest absolute AUC ROC of 0.707. While both subjects boast an inordinately high imbalance (>90\%) in their ground truth responses--- subject S30 responding ``not seen" for 1,828 videos out of 2,000, and subject S36 doing so for 1,871 videos---subject S30's results are likely a reflection of the quality of their response data as a large portion of their response reaction times consistently rested within the 1 - 1.5 second mark. This suggests a more rhythmic rather than innate and earnest nature to their responses, whereas subject S36's average response reaction times were more varied and typically given <1 second after stimulus onset.

Figure ~\ref{fig:rain2} shows the per-subject distributional differences in video ground-truth memorability (population level) scores for remembered (tp) and forgotten (fn) videos. A series of statistical tests were carried out in order to assess the relationship between response reaction times and memorability scores.

\subsection{Statistical Analysis}

First, we conducted a Pearson correlation analysis between response times and ground-truth memorability scores for each video across all subjects. Our findings revealed no statistically significant correlations between these variables for any subject, suggesting that no clear linear relationship exists between response times and memorability scores. Then we performed two per subject independent t-tests with Bonferroni correction, testing the differences between mean response reaction times and mean video memorability scores for remembered and forgotten videos. For mean response reaction times, we found significant differences for subjects 10, 13, 16, 19, 30, 36, 37, 9, and 2, where the mean response reaction times for remembered videos were consistently higher. For subjects 41 and 4, we found the inverse. Subject 31 was the only subject with no statistically significant difference in their mean response reaction times. For ground-truth memorability scores, we found that only subjects 2 and 4's results showed significant differences in mean memorability scores. For subject 2, the mean memorability score was higher for remembered videos (M=0.842) compared to not remembered videos (M=0.771), t=6.801, p=3.33e-11. Likewise, for subject 4, the mean memorability score was higher for remembered videos (M=0.809) compared to not remembered videos (M=0.780), t=2.891, p=0.00395. Given that only subjects 2 and 4 demonstrated a significant difference, it highlights the fact that the influence of memorability on long-term video recognition might be inconsistent across individuals, or potentially affected by individual differences or data quality variations. Further investigation is required to determine if there is a causal relationship or if other factors might be influencing the observed differences.

\section{Conclusion}

The results of our research support both the hypothesis that theta band (4--8Hz) oscillations over the right temporal lobe (channel P8) are involved in the encoding of declarative memory, and the potential existence of a distinct moment of memorability that can be leveraged to predict subsequent subject-specific recognition. Although we did not find a significant interaction between subject-dependent (SD) and subject-independent (SI) training approaches, the better performance of several SD trained models suggests the potential importance of subject-specific EEG data for prediction. This highlights the need for further investigation into the adaptability and generalisability of models across different individuals. 

One limitation of the current study is the reliance on binary ground truth responses, which may not fully capture the intricacies of memory formation and recall. Additionally, the presence of volume conduction in EEG measures warrants further exploration of methods to mitigate or account for its impact on the results. While the study has unveiled promising avenues for further exploration, demonstrating the feasibility of using EEG signals and deep learning techniques to predict video memorability, it also raises new questions and challenges for future research.

\begin{acks}
This publication has emanated from research partly supported by Science Foundation Ireland (SFI) under Grant Number SFI/12/RC/ 2289\_P2 (Insight SFI Research Centre for Data Analytics).
\end{acks}

\bibliographystyle{ACM-Reference-Format}
\bibliography{bib}
\end{document}